# Ab-initio cluster approach for high-harmonic generation in liquids

Ofer Neufeld[1,*], Zahra Nourbakhsh[1], Nicolas Tancogne-Dejean[1], Angel Rubio[1,2]

[1]Max Planck Institute for the Structure and Dynamics of Matter and Center for Free-Electron Laser Science, Hamburg, 22761, Germany.
[2]Center for Computational Quantum Physics (CCQ), The Flatiron Institute, New York, NY, 10010, USA.

**ABSTRACT:** High harmonic generation (HHG) takes place in all phases of matter. In gaseous atomic and molecular media, it has been extensively studied and is very well-understood. In solids research is ongoing, but a consensus is forming for the dominant microscopic HHG mechanisms. In liquids on the other hand, no established theory yet exit and approaches developed for gases and solids are generally inapplicable, hindering our current understanding. We develop here a powerful and reliable *ab-initio* cluster-based approach for describing the nonlinear interactions between isotropic bulk liquids and intense laser pulses. The scheme is based on time-dependent density functional theory and utilizes several approximations that make it feasible yet accurate in realistic systems. We demonstrate our approach with HHG calculations in water, ammonia, and methane liquids, and compare the characteristic response of polar and non-polar liquids. We identify unique features in the HHG spectra of liquid methane that could be utilized for ultrafast spectroscopy of its chemical and physical properties: (i) a structural minima at 15-17eV, and (ii) a well-like shape in the perturbative region that is reminiscent of a shape resonance. Our results pave the way to accessible calculations of HHG in liquids and illustrate the unique nonlinear nature of liquid systems.

## 1. INTRODUCTION

High harmonic generation (HHG) is an extremely nonlinear optical process that occurs when intense laser fields are irradiated onto material media. Interactions between electrons in the medium and the incident laser result in an up-conversion of photons, emitting a spectrally-wide frequency comb that reaches up to XUV energies[1]. HHG has been experimentally demonstrated in all phases of matter, namely in gases, solids, and liquids. In the gas phase, where it was first discovered[2,3], HHG has been extensively researched for several decades and is very well-understood. Here it is commonly described by a semi-classical[4–6] (or quantum[7,8]) three-step model that intuitively describes the process by three sequential steps: (i) ionization of an electron due to laser-induced suppression of the binding coulomb potential, (ii) acceleration of the liberated electron in the continuum whereby it gains kinetic energy, and (iii) a recombination of the liberated electron with its parent ion (or ions in molecules) that results in the emission of high energy photons. This model is routinely used to explain experimental results and to develop new spectroscopy and interferometry approaches. Notably, the three-step model relies on the fact that the atoms or molecules in the gas are isolated in real-space.

In recent years it was shown that solids are also a prominent source of high harmonics with some possible advantages over their gas-driven counterparts[9,10]. The mechanism for HHG in solids differs from that in gases, and mainly relies on the interference from two types of emissions: (i) emission due to intraband motion of electrons within the non-parabolic band structure, and (ii) interband emission due to electron-hole recombination, which is analogous to the gas-phase three-step model except that electrons accelerate along the bands in *k*-space[11–19]. Notably, the theory that describes these mechanisms relies on long-range translational symmetries of the solid (i.e. a band-structure picture). It is important to point out that fundamental understanding of the HHG process in both gases and solids is the driving force behind technologies and applications based on high harmonics, e.g. attosecond pulses[1,20,21], novel ultrafast spectroscopies[22–29], and imaging techniques[30–34]. From a practical standpoint, developing new ultrafast (and potentially attosecond) spectroscopies based on HHG is highly appealing, since the spectra is usually extremely sensitive for any internal structure in the nonlinear media (e.g. its symmetry[35–38], topology[29,39–43], chirality[44–48], etc.) due to the highly nonlinear nature. The prospects of transferring some of the ideas and methods implemented in gases and solids to the liquid phase is exciting, since most biochemical processes occur in liquid or hydrated phases.

Contrary to gases and solids, liquid HHG measurements are quite scarce. HHG was observed in liquid microdroplets[49] and surfaces[50], and in the perturbative regime from bulk liquids[51]. More recently, XUV high harmonics were measured and characterized from bulk liquids[52]. Here there are many fundamental open questions. For instance, the HHG cutoff scaling law is still under debate[52], the dominant generation mechanisms have yet to be uncovered, it is not yet clear how the process depends on the physical and chemical properties of the liquid, and the list goes on. A single theoretical work attempted to answer some of these questions[53], but it relied on a one-dimensional toy model of a defected solid rather than a realistic liquid. Indeed, the main challenges for a description of liquids is that one usually needs large molecular ensembles in order to correctly capture their short-range coordination and isotropic nature, while many hybridized molecular states are chemically and optically active. Other effects such as hydrogen bond dynamics are also notoriously difficult to simulate[54–56]. At the same time, mechanisms and analytic approaches developed for gaseous and solid media are inapplicable to liquids, because they lack long-range correlations, and are dense infinite systems. Nevertheless, a proper and feasible description of strong light-matter interactions in liquids is necessary to answer all of the fundamental questions above, and the lack of such an approach is one of the main reasons that liquid HHG has been poorly understood thus far.

Here, we develop an *ab-initio* approach for strong light-matter interactions in realistic liquids interacting with arbitrarily polarized laser pulses. The scheme is based on time-dependent density functional theory (TDDFT) for large molecular clusters, although it can also be implemented with other *ab-initio* techniques. In order to make calculations feasible, we utilize several approximations for the dynamics: (i) dynamical electron-electron correlations are frozen in time, (ii) contributions of surface-localized states to the nonlinear response are suppressed, (iii) contributions of deep-lying states to the nonlinear response are neglected, and (iv) the cluster response is minimally orientation averaged to mimic an isotropic system. These approximations are tested explicitly, and also by comparing to experimental results[57]. The model is then employed for HHG calculations in liquid water, ammonia, and methane. We compare the characteristic response of polar and non-polar liquids, and find that non-polar liquids lead to much sharper harmonic peaks with suppressed interference effects. We show that the HHG spectra from liquid methane contains interesting features that could be used for ultrafast spectroscopy, including a structural-minima at 15-17eV, and a well-like shape in the perturbative region.

The paper is organized as follows. In section 2 we introduce our approach and the logic behind it. In section 3 we analyze HHG in liquid water in various laser conditions. Section 4 addresses the main differences between HHG from polar and non-polar liquids. Finally, section 5 summarizes our results and presents an outlook.

## 2. METHOD FORMULATION

We begin with a formal description of our approach. The liquid is described with relatively large molecular clusters of 40-60 molecules. The geometries of the clusters can be readily obtained as minimal energy configurations[58,59], or from molecular dynamics simulations[60–62] (see illustration in Fig. 1(a) for minimal energy configuration of a water cluster). The ground state of each cluster is obtained using real-space grid-based DFT calculations with octopus code[63–66]. The real-space formulation allows us to employ a minimal spherical box shape that has additional vacuum spacing, where the vacuum layer size is defined as the distance between the outermost atom in the cluster and the box wall. In this paper we consider PBE exchange-correlation (XC)[67] with an added van-der-waals correction[68] (though the particular XC choice can vary according to the studied system), and core states are replaced by norm-conserving pseudopotentials[69]. We neglect here the spin degree of freedom for simplicity. Additional technical details are delegated to the Supplementary Information (SI).

Upon obtaining the Kohn-Sham (KS) orbitals that comprise the ground state electron density, we analyze their structure to identify any surface-localized bands as opposed to de-localized states (see illustration in Fig. 1(b) and (c)). This step is crucial, since the large surface to volume ratio of clusters often leads to significant localization even for closed-shell molecules. However, our goal is to describe the response of bulk liquids. Accordingly, surface-localized states are cataloged such that their response in the time-dependent calculations can be removed. Here we remove the time-dependent response of last band of occupied orbitals in polar liquids, which are largely surface-localized (but results are mostly insensitive to the inclusion/exclusion of a small number of orbitals due to the system size). We note that this choice is not unique, and in the future one may derive a more rigorous mathematical condition to determine surface locality. It is also noteworthy that surface-localization is the result of bonding between molecules, e.g. through hydrogen bonds, van-der-waals interactions, or other sources. Consequently, weakly-bonded liquids (e.g. liquid Helium) do not require this procedure.

In the next step, we wish to describe the interaction of the liquid with an incident laser pulse. This is accomplished within TDDFT, where the KS orbitals are propagated with the following coupled equations of motion (we use atomic units throughout):

$$i\partial_t |\varphi_j^{KS}(t)\rangle = \hat{h}(t)|\varphi_j^{KS}(t)\rangle \qquad (1)$$

where $|\varphi_j^{KS}(t)\rangle$ is the $j$'th time-dependent KS state and $\hat{h}(t)$ is the one-body Hamiltonian:

$$\hat{h}(t) = -\frac{1}{2}\nabla^2 + v_{KS}(\mathbf{r},t) - \mathbf{E}(t)\cdot\mathbf{r} \qquad (2)$$

, and where $v_{KS}(\mathbf{r},t)$ is the time-dependent KS potential that is given in the adiabatic approximation by:

$$v_{KS}(\mathbf{r},t) = -\sum_I \frac{Z_I}{|\mathbf{R}_I - \mathbf{r}|} + \int d^3r' \frac{\rho(\mathbf{r}',t)}{|\mathbf{r}-\mathbf{r}'|} + v_{XC}[\rho(\mathbf{r},t)] \qquad (3)$$

Here $Z_I$ is the charge of the $I$'th nuclei in the cluster and $\mathbf{R}_I$ is its coordinate, $v_{XC}$ is the XC potential that is a functional of $\rho(\mathbf{r},t)=\sum_j|\langle r|\varphi_j^{KS}(t)\rangle|^2$, the time-dependent electron density. The motion of the nuclei is neglected, which is justified for interactions with ultrashort laser pulses (even in longer pulses effects are expected to be small[70–72]). Note that the bare coulomb interactions of electrons with the nuclei in eq. (3) is replaced by pseudopotentials in calculations in order to reduce computational costs. $\mathbf{E}(t)$ in eq. (2) is the electric field vector of a laser pulse with an arbitrary polarization and carrier frequency. We use the dipole approximation and neglect the spatial dependence of the electric field, which is justified for laser wavelengths that are much larger than the cluster sizes. Accordingly, we also neglect interactions with the magnetic field components of the laser, and any other relativistic terms such as spin-orbit coupling (these can be added in a straightforward manner). We use the length gauge for describing the light-matter interaction term, but equivalent forms can also be utilized. Lastly, we note that the initial KS orbitals are taken as their ground state forms.

Before solving these equations, we must address several points: (i) Remove any surface and finite-size effects from the response in order to capture only bulk contributions. (ii) Recall that unlike the bulk liquid, the cluster is not perfectly isotropic (this is true even for large clusters). We address point (i) by freezing the occupation of the localized surface states to their initial value, i.e. $|\varphi_s^{KS}(t)\rangle = |\varphi_s^{KS}(t=0)\rangle$ for any '$s$' that corresponds to surface-localized states. This guarantees that inner molecules in the cluster feel the correct mean-field potential that is still affected by surrounding electrons, as they would in the bulk (because the outer-shell molecules do not get ionized). Furthermore, the surface states themselves do not contribute to the response of the liquid because they are kept static. Practically, this means that the total electron density, $\rho(\mathbf{r},t)$, is divided to a dynamical piece that is allowed to evolve, $\rho_{dyn}(\mathbf{r},t)$, and a frozen piece, $\rho_{frz}(\mathbf{r},t)$ that corresponds to a static charge density. We also suppress any additional surface response by adding a complex absorbing potential (CAP) to the vacuum region (see illustration in Fig. 1(a), and details in the SI)[73]. This means that a non-Hermitian term, $v_{CAP}(\mathbf{r})$, is added to the total KS potential in eq. (2) during temporal evolution. Point (ii) is addressed by performing a minimal orientation averaging of the cluster's response through trapezoidal weights (see SI for details), i.e. one must perform several calculations with the laser polarization axis rotated in three-dimensional space. These procedures are motivated by the assumption that the liquid response should correspond to that of the inner caged molecules in the cluster, since those feel the correct bonding with neighboring molecules and have the proper short-range symmetry and coordination.

The orientation averaging is meant to mimic the isotropic response of a much larger liquid volume that is inaccessible in calculations such that the laser 'sees' many inter-molecular configurations that are summed over.

At this point we note that even after having performed the approximations above, solving the set of coupled TDDFT KS equations for the cluster is a challenging task. For instance, for a modest cluster size of 50 molecules where each molecule contributes just four active states, there are 200 active orbitals that need to be propagated in tandem, self-consistently, and on large real-space grids. This needs to be performed consecutively for several laser orientations, and for a reasonably long simulation time (in order to obtain spectrally-resolved harmonics). To make calculations more accessible, we employ additional approximations. First, we freeze the KS potential to its ground state initial form, i.e. $v_{KS}(\mathbf{r}, t) = v_{KS}(\mathbf{r}, t = 0)$, which fully uncouples the equations of motion for the KS orbitals. This approximation is valid only for relatively moderate laser powers where $\rho(\mathbf{r}, t)$ does not change drastically. It is the equivalent of the non-interacting electrons approximation that has seen great success in both gas and solid HHG[1,18,19,21,74]. We test this approximation and make sure that it is valid in the SI. Second, we freeze the response of any deeper-lying states that contribute negligibly to the optical response. This is analogous to the single-active-electron approximation that is standardly used in gas phase HHG[1], but where only the lowest energy band of states is frozen (it is also analogous to limited-band models in solid HHG[19,21]). Altogether, we are left with uncoupled equations of motion for the remaining orbitals (those that are not deep-lying, nor surface localized), which constitutes a significant reduction in the problem size.

Upon propagating the KS orbitals (see SI for numerical details) we obtain $\rho(\mathbf{r}, t)$, from which the induced microscopic polarization is given as:

$$\mathbf{P}_\alpha(t) = \int d^3r \, \mathbf{r} \rho(\mathbf{r}, t) \tag{4}$$

where $\alpha$ denotes the solid-angle orientation of the cluster with respect to the laser. Following orientation averaging, the induced polarization of the isotropic liquid is given as:

$$\mathbf{P}(t) = \int d\alpha \, \mathbf{P}_\alpha(t) \tag{5}$$

The dipole acceleration $\mathbf{a}(t)$, is found directly by the second temporal derivative of $\mathbf{P}(t)$. The harmonic spectrum is given by the Fourier transform of $\mathbf{a}(t)$: $I(\Omega) = \left| \int dt \, \mathbf{a}(t) e^{-i\Omega t} \right|^2$.

It is helpful to briefly summarize the numerical parameters of the approach to be converged. First, there are the standard parameters for the DFT calculations of the ground state, e.g. spacing and grid dimensions. Second, there are numerical parameters of the propagation scheme, including time-steps, the vacuum spacing, and parameters of the CAP. Lastly, there are the conceptual details of the model: the cluster size and the angular orientation grid density. Convergence data is presented in the SI.

As a final note on the feasibility and accessibility of the approach, we highlight the order of magnitude of the required resources – we obtain a single converged HHG spectra for a linearly-polarized laser (at 800nm wavelength with 8 optical cycle long pulses (21.3 fs)) from clusters with ~50 molecules (~100 active KS states) in ~15,000 CPU hours. Parallelized over 256 CPUs, this is ~2 days per spectra. These figures are comparable in magnitude to those required from TDDFT calculations for solid HHG (depending on the system).

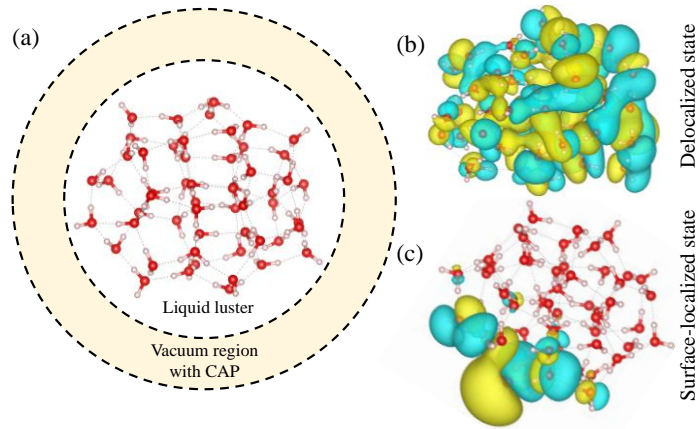

**Figure 1.** Conceptual illustration of the proposed cluster approach to study HHG in liquids. (a) An approximately spherical cluster with a geometry and density that corresponds to a liquid phase is embedded in a real-space grid with a spherical boundary. The cluster is encapsulated by an absorbing layer that passivates the nonlinear response associated with the surface to mimic a bulk liquid. This illustration depicts a liquid water cluster with 54 $H_2O$ molecules obtained from ref. 59. (b) The exemplary 151$^{st}$ KS state that is delocalized and is part of several bands of delocalized states. (c) Same as (b), but for the 216$^{th}$ KS states that is highly surface-localized, and is part of a fully localized topmost band of states.

## 3. RESULTS AND DISCUSSION

**3.1. HHG in liquid water.** Having outlined our approach, we now utilize it to perform HHG calculations in liquids. For simplicity, we explore monochromatic linearly-polarized laser pulses of the form:

$$\mathbf{E}(t) = f(t)E_0 \cos(\omega t)\hat{x} \qquad (6)$$

where $\omega$ is the fundamental frequency, $E_0$ is the field amplitude, and $f(t)$ is a trapezoidal envelope function with two-cycle long rise and drop sections, and a four-cycle long flat top section.

We begin by analyzing HHG in liquid water. Cluster geometries were obtained from ref. 59 as minimal energy configuration of the AMOEBA force-field approach[75] (see Fig. 1(a)). In water, we find that the highest energy hybridized band of orbitals is largely surface-localized (see illustration in Fig. 1(c)). The deepest band of orbitals contributes negligibly to the HHG response (see SI). Thus, there are two active bands comprised of $N$ orbitals each ($N$ being the number of molecules in the cluster). Fig. 2 presents exemplary HHG spectra obtained at various laser wavelengths and powers. It is immediately apparent that the spectra in Fig. 2 contain only odd harmonics (indicated by dashed gray lines in Fig. 2). We further note that all harmonics have only $x$-polarized components. These fundamental symmetry constraints[38,76] indicate that the nonlinear optical response of the cluster is indeed isotropic, as required from a bulk liquid. It also suggests that the surface response is correctly suppressed, because it would result in non-isotropicity.

Figures 2(a,b) also compare these results to HHG calculations from a single gas-phase isolated water molecule (calculated on a similar level of theory, see SI for details). The HHG cutoff from the single-molecule case considerably deviates from the cutoff in the liquid, hinting towards the different mechanisms active in each system (dilute gases or liquids). Notably, there is some emission noise beyond the cutoff of the liquid that corresponds to energy ranges where harmonics are still emitted from the gas (e.g. see noisy emission beyond 35 eV in Fig. 2(b)). This noise can be interpreted as HHG contributions from electrons that are ionized from the cluster, but are adequately absorbed by the CAP. Overall, we conclude that the approach indeed manages to suppress surface contributions and to mimic the isotropicity of a bulk liquid. Importantly, these results hold for all examined laser parameters (wavelengths of 800-1500nm and intensities $3-8\times10^{13}$ W/cm$^2$) and liquids, supporting the generality of the technique.

At this point we highlight that the numerical technique developed here was recently used to explore the cutoff scaling of liquid HHG with respect to wavelength, and to derive an intuitive picture for the HHG mechanism in liquids[57]. Ref. 57 has demonstrated that the cluster approach successfully reconstructs the experimentally measured wavelength-independent cutoff (contrary to the standard behavior in gases and

solids[1,21]). This further establishes the validity of the model and its utilization for exploring fundamental phenomena in liquids.

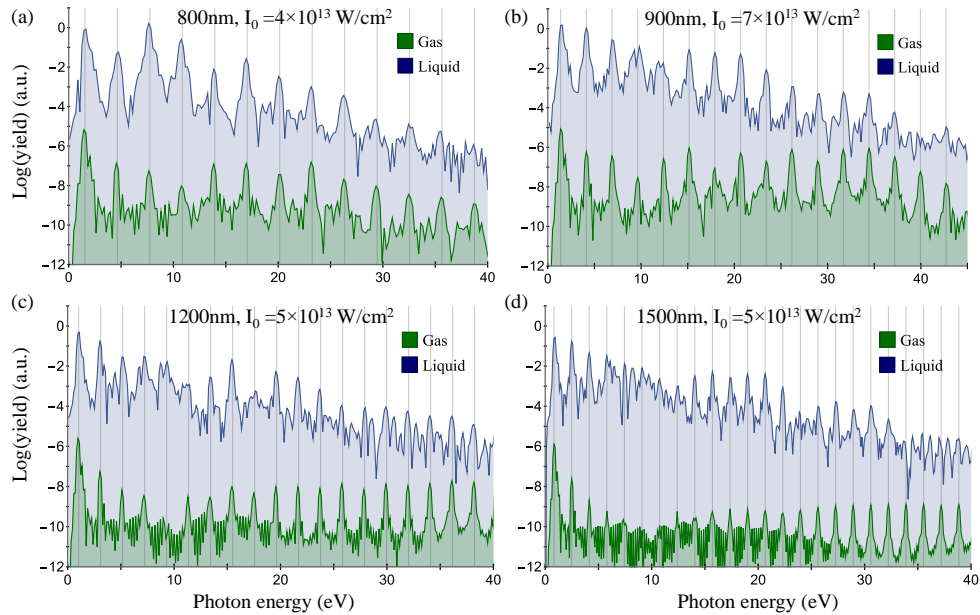

**Figure 2.** HHG spectra calculated with the cluster approach for liquid water (blue) at various laser conditions (a-d). Green spectra represent calculations for the single-molecule gas-phase case in similar settings (these have been artificially reduced in power to enhance visibility). Gray lines denote positions of odd harmonics.

**3.2. HHG from polar and non-polar liquids.** We next explore HHG from two additional liquids: ammonia and methane. Liquid methane cluster geometries are obtained from ref. 58 as minimal energy configurations based on *ab-initio* obtained potentials[77], while liquid ammonia geometries are taken from ref. 61 utilizing a molecular dynamics approach. We repeat the calculations performed in the previous part for both of these liquids and scan various laser wavelengths and powers. Figures 3 and 4 present results in similar nature to those seen in water: only *x*-polarized odd harmonics are emitted indicating that the response is isotropic. We also note that a converged isotropic response for these molecules is obtained for less orientations compared to the case of $H_2O$ as a result of their higher symmetry (see SI for details).

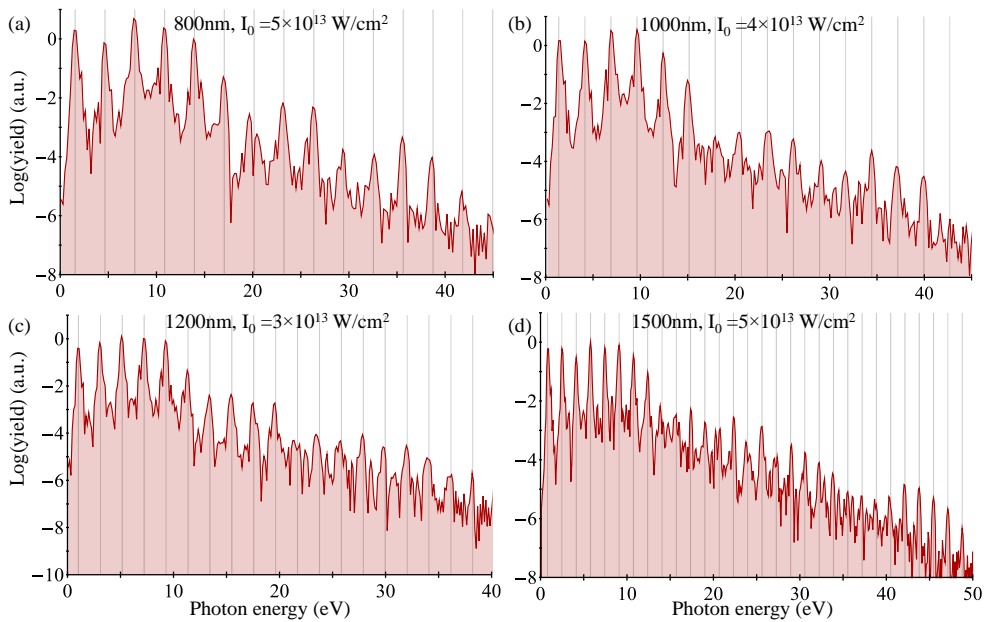

**Figure 3.** HHG spectra calculated with the cluster approach for liquid ammonia at various laser conditions (a-d). Gray lines denote positions of odd harmonics.

It is worthwhile to examine the characteristic differences between the nonlinear response of these various liquids. Most notably, both ammonia and water exhibit strong inter-molecular bonding that arises from van-der-waals interactions and hydrogen bonding, unlike in methane where each molecule is nearly non-bonded to its neighbors. This fundamental difference also means that methane has practically no surface-state localization. Comparing the HHG spectra in Figs. 2 and 3 to those in Fig. 4, the spectra from methane comprises of much sharper distinct harmonic peaks as opposed to water and ammonia. In water for instance, some harmonics show sideband oscillations (e.g. the harmonic at 30eV in Fig. 2(a)), or tend to split into sub-peaks (e.g. the harmonic at 10eV in Fig. 2(b)). The same effect occurs in ammonia (see for example the harmonics at 20eV and 25eV in Fig. 3(a)). This is likely a result of multi-orbital interference in water and ammonia liquids that can originate from inter-molecular recombination or scattering. These interference effects can also be understood to arise in the bonded liquids because they exhibit wide energy bands that allow intricate coupled inter-band and intraband dynamics in *k*-space. For methane however, this effect does not occur. Its absence suggests that liquid methane exhibits a more dominant single-molecule response. In fact, the 'cleanness' of the harmonic peaks in liquid methane is reminiscent of spectra usually obtained from gas phase calculations of isolated molecules. We attribute this to the non-bonding nature of the non-polar liquid that suppresses inter-molecular interferences (in *k*-space this can be thought of as arising from the highly energy-resolved bands that have a uniform character which suppresses interband-intraband interferences). This feature might be useful in future studies for probing hydrogen bonding dynamics during chemical processes.

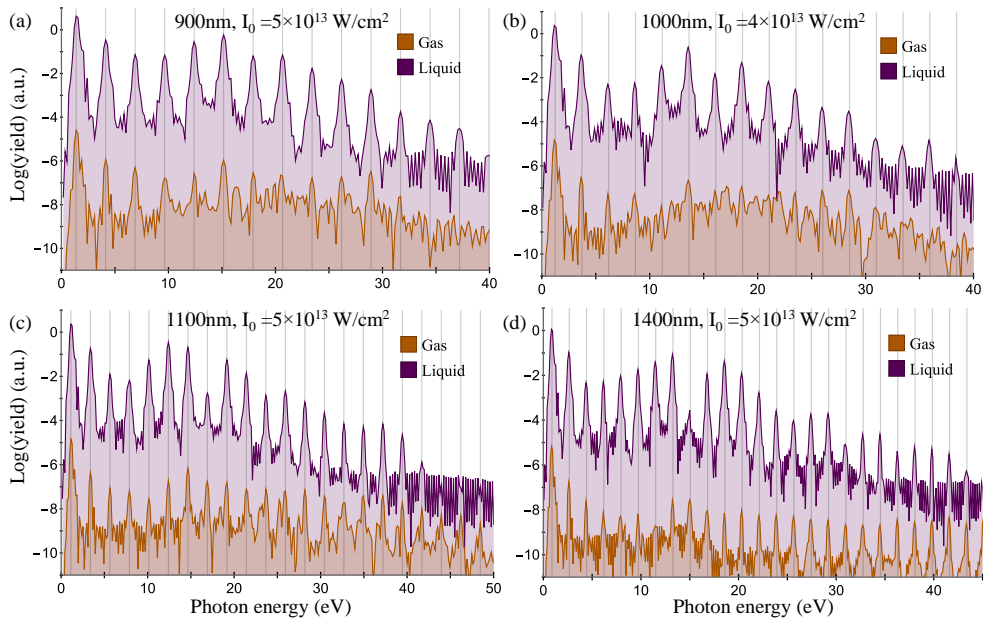

**Figure 4.** HHG spectra calculated with the cluster approach for liquid methane (purple) at various laser conditions (a-d). Orange spectra represent calculations for the single-molecule gas-phase case in similar settings (these have been artificially reduced in power to enhance visibility). Gray lines denote positions of odd harmonics.

The exceptionally clean HHG spectra from methane can pose a unique advantage for exploring its structural and electronic properties on ultrafast timescales. This is because it may be more sensitive to small interference effects (whereas in polar liquids these are more difficult to disentangle). In particular, we note two interesting features that arise in the liquid HHG emission from methane that could be utilized for this purpose:

(i) The harmonic emission at energy ranges up to 14 eV exhibits a very distinct well-like shape with a typical minima around 6eV. That is, the envelope of the harmonic spectra in this energy region has a unique behavior – the yield exponentially drops, reaches a minimum, and then exponentially increases. This is a different behavior than that usually observed in both gases and solids, where the perturbative region shows a simple exponential decay. Figure 5(a,b) presents the integrated harmonic power from methane liquid in this energy range for various laser parameters - the well shape is observed for a wide regime of laser conditions with wavelengths ranging from 800-1500nm, and laser powers of $3-8\times10^{13}$ W/cm$^2$ (though the specific shape of the well slightly varies with the parameters). We note that a similar effect is observed from methane gas, but it is much weaker compared to the liquid case (see Fig. 4). This could indicate that the well originates from the chemical properties of the single CH$_4$ molecule that is dressed by the surrounding environment. This phenomenon is not observed in the other tested liquids (see Figs 2-3). The particular shape of the well is reminiscent of that resulting from shape resonances in barrier-well systems, as was demonstrated in 1D models[78]. Thus, HHG in liquids could pave the way to novel spectroscopic probing of resonances in the electronic structure.

(ii) The HHG spectra from methane shows a clear structural interference minima at 15-17 eV (see Fig. 4 and 5(c)). The emission at this energy range remains at a stable local minima even when changing the laser wavelength for a wide range of 900-1500nm, and for laser powers of up to $5\times10^{13}$ W/cm$^2$. Furthermore, the minima is not present in gas phase spectra. Thus, we conclude that it is associated solely with the chemical and physical properties of the liquid system. Notably, the minima is washed-out at stronger laser powers, possibly indicating that different mechanisms or pathways become dominant. We also note that it is slightly less pronounced when including dynamical correlations in the calculation, though still visible (see SI). This should be investigated in future work. We emphasize that this is the first prediction of a structural minima in the HHG spectra of a liquid system, which is equivalent to those seen in gases[79–81] and solids[82], and which can potentially be used to probe correlations or other dynamical effects.

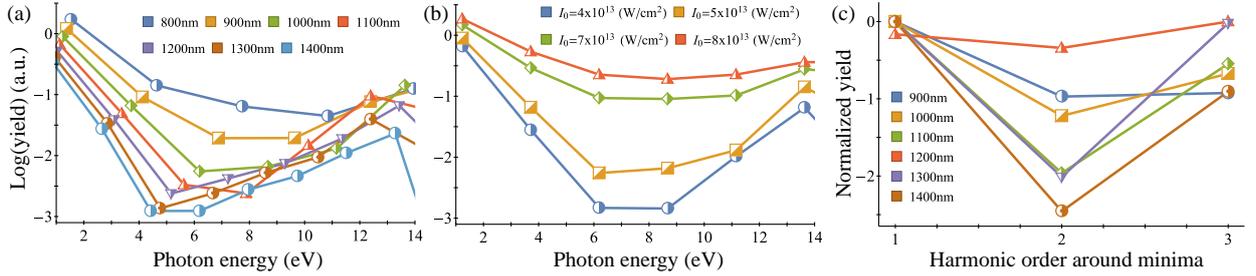

**Figure 5.** Wavelength and intensity dependent analysis of well-shape minima and structural interference in HHG from liquid methane. (a) Integrated harmonic yield per harmonic order in the perturbative region for various laser wavelength (calculated for $I_0=5\times10^{13}$ W/cm$^2$). (b) Same as (a) but for various laser powers (calculated at λ=1000nm). (c) Integrated harmonic yield for varying laser wavelengths in the region of a structural interference minima (15-17eV), calculated at $I_0=5\times10^{13}$ W/cm$^2$. Harmonic yield is presented for three harmonic orders around the minima in each case (the minima is shifted to harmonic #2), and the maximal power is normalized to 0.

## 4. CONCLUSIONS AND OUTLOOK

To summarize, we have put forward a cluster-based *ab-initio* approach for describing interactions between bulk liquids and arbitrarily-polarized intense laser pulses. This technique formally relies on TDDFT, but utilizes several approximations that allow affordable and accurate calculations for realistic three-dimensional systems. It has also been validated by agreement with experiments[57]. We implemented our technique to study HHG from liquid water, ammonia, and methane, and investigated the role of the liquid's chemical properties on the spectra. We concluded that spectra from non-polar weakly-bonded liquids shows much sharper harmonic peaks than those from polar liquids that exhibit stronger interference effects. We have also shown that the HHG spectra from liquid methane exhibits some interesting characteristics that may be useful for ultrafast spectroscopy: (i) a local minima in the perturbative harmonic region that might originate from shape resonances[78]. (ii) An electronic-structure minima that appears at 15-17 eV, which is the equivalent to those observed in gases[79–81] and solids[82]. Both of these features could be utilized to study the various chemical and physical properties of liquids with ultrafast temporal resolution (e.g. dynamical correlations, dynamical polarizability, ion motion, etc.).

Apart from the specific predictions presented here, we believe that our approach might pave the way for feasible and accessible calculations of HHG in liquids, as well as other nonlinear processes including photoionization[83]. This is a crucial step towards improved understanding of the active mechanisms for strong-field physics in liquids, which is essential for obtaining novel light sources and ultrafast spectroscopic capabilities.

- **ASSOCIATED CONTENT**

**Supporting information**

Supporting information is available for this article, including technical details of the methodology, convergence tests, and validation of several of the approximations used.

- **AUTHOR INFORMATION**

**Corresponding Authors**

Ofer Neufeld – Max Planck Institute for the Structure and Dynamics of Matter and Center for Free-Electron Laser Science, Hamburg 22761, Germany; ofer.neufeld@mpsd.mpd.de

Anegl Rubio – Max Planck Institute for the Structure and Dynamics of Matter and Center for Free-Electron Laser Science, Hamburg 22761, Germany; angel.rubio@mpsd.mpg.de

**Notes**

The authors declare no competing financial interests.

- **ACKNOWLEDGMENTS**

We thank Hans Jackob Wörner, Mondal Agana, Zhong Yin, and Vit Svoboda, for helpful discussions. We thank Esam A. Orabi and Guillaume Lamoureux, and Zhong-Zhi Yang and Dong-Xia Zhao for providing us with geometries for liquid ammonia clusters. We acknowledge financial support from the European Research Council (ERC-2015-AdG-694097). The Flatiron Institute is a division of the Simons Foundation. This work was supported by the Cluster of Excellence Advanced Imaging of Matter (AIM), Grupos Consolidados (IT1249-19) and SFB925. O.N. gratefully acknowledges support from the Alexander von Humboldt Foundation and from a Schmidt Science Fellowship.

- **REFERENCES**

# Supplementary information: Ab-initio cluster approach for high harmonic generation in liquids

Ofer Neufeld[1,*], Zahra Nourbakhsh[1], Nicolas Tancogne-Dejean[1], Angel Rubio[1,2]

[1]Max Planck Institute for the Structure and Dynamics of Matter and Center for Free-Electron Laser Science, Hamburg, 22761, Germany.
[2]Center for Computational Quantum Physics (CCQ), The Flatiron Institute, New York, NY, 10010, USA.

This supplementary information (SI) file contains technical details on the numerical calculations and methodology used in the main text, as well as some additional results. Section S1 presents details of the numerics, including convergence testing. Section S2 presents some additional results of liquid HHG and tests for the validity of various approximations used in the main text.

- **S1: NUMERICAL DETAILS**

### 1. Ground state DFT calculations

All DFT calculations were performed using the octopus code[1–3]. The Kohn-Sham (KS) equations were discretized on a Cartesian grid with a spherical shape of radius 31.6, 34, and 32.8 bohr for the converged clusters of water, ammonia, and methane, respectively, and 45 bohr for the single-molecule calculations. Calculations were performed using the PBE exchange correlation functional in all cases[4]. For the liquid clusters, a van-der-waals correction term was added[5]. For the single molecule cases, a self-interaction correction (SIC) was added in order to capture the long-range asymptotic potential[6]. The frozen core approximation was used for core states, which were treated with norm-conserving pseudopotentials[7]. The KS equations were solved to self-consistency with a tolerance $<10^{-7}$ Hartree, and the grid spacing was converged to $\Delta x=\Delta y=\Delta z=0.4$ bohr, such that the total energy per electron was converged $<10^{-3}$ Hartree. Cluster geometries were obtained as described in the main text. Single molecule geometries were taken at the experimental configuration.

### 2. Time-dependent calculations

For time-dependent calculations, the active KS orbitals were propagated with a time step $\Delta t=0.11$ a.u. and by adding a complex absorbing potential (CAP) with a width of 11.2 bohr in the vacuum region[8]. The initial state was taken to be the system's ground-state. The grid size, absorbing potential, and time step were tested for convergence. The HHG spectra were obtained as explained in the main text where the dipole acceleration was filtered with a super-gaussian window.

### 3. Orientation averaging

Orientation averaging was performed using trapezoidal weights with an angular grid spanned by Euler angles in the *z-y-z* convention. For the liquids we considered three angular grids with Euler angle spacing of: (1) $\pi/2$ for all angles, (2) of $\pi/2$ for $\alpha$ and $\gamma$ and $\pi/4$ for $\beta$, and (3) $\pi/4$ for all angles. After utilizing the symmetries of linearly-polarized light in the dipole approximation, these grids lead to a total of 6, 14, and 58 independent orientations, respectively (which are equivalent to full grids with 75, 125, and 405 orientations, respectively). The HHG spectra in liquid water were converged with grid (2). The HHG spectra of ammonia and liquid methane (which have a higher molecular symmetry than water) were converged with grid (1). For gas-phase calculations we used grid (3).

### 4. Cluster size convergence

Convergence with respect to cluster size was obtained for a cluster of 54 molecules in the case of water (see Fig. S1(a) showing small disparities for harmonics near 15 and 30eV, but a generally similar spectral structure). For the case of ammonia convergence was obtained for a 56-molecule cluster (see Fig. S1(b) showing small deviations above 30 eV). In liquid methane convergence was obtained for a 40-molecule cluster (see Fig. S1(c) showing small deviations only above 30 eV). In all cases there are small deviations upon increasing cluster size by 10-15%, even though the molecular geometries in each case can be very different (because they were obtained as minimal energy configurations of completely different clusters). This is a good indication for the reliability of the method, and also suggests that the surface contribution is indeed suppressed (because surface reconstruction variations are negligible). Notably, higher energy harmonics are slightly more difficult to converge. To compensate for this effect and reduce errors even further we use the larger cluster sizes in all calculations.

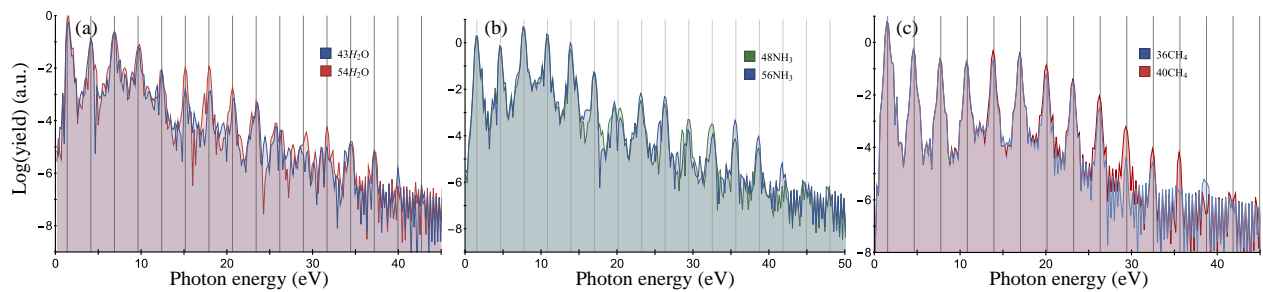

**Figure S1.** Convergence with respect to cluster size. (a) Liquid water HHG spectra calculated at $\lambda$=900nm and a laser power of $4\times10^{13}$ W/cm$^2$. (b) Liquid ammonia HHG spectra calculated at $\lambda$=800nm and a laser power of $5\times10^{13}$ W/cm$^2$. (c) Liquid methane HHG spectra calculated at $\lambda$=800nm and a laser power of $5\times10^{13}$ W/cm$^2$. Gray lines indicate the position of odd harmonics.

## 5. Gas-phase calculations

For time-dependent gas phase calculations we utilized exactly the same approach as in the cluster cases. The only differences are that: (1) we performed orientation averaging with a much denser angular grid (see discussion above), and (2) employed a SIC correction rather than a wan-der-waals correction (see discussion above). The same level of theory was used in order to have comparable results between the clusters and the single-molecule calculations, i.e. PBE XC was used throughout and the KS potential was frozen to its initial form.

- **S2: ADDITIONAL RESULTS**

### 1. Validity of the non-interacting electrons approximation

Throughout the paper we utilized the non-interacting electrons approximation, i.e. the KS potential was frozen to its ground state form. This approximation is standardly used in both gas and solid phases, but has never before been tested for liquids. Here we formally test it for the case of liquid methane (which is simpler due to the lack of surface localization). Figure S2(a) presents HHG spectra from liquid methane obtained at similar conditions where the KS potential is either kept frozen in time, or allowed to evolve (a standard TDDFT calculation). Some differences indeed emerge between the spectra such as differences in the total harmonic power and small a cutoff value shift. Nonetheless, the envelope structure of the spectra and its characteristic shape is largely unchanged, including for instance the position of the observed interference minima at ~15-17 eV (see main text). We further argue that the source of the disparity between the two approaches is an unphysical renormalization of the KS potential due to spurious ionization from the cluster that is absorbed at the boundaries (as is seen in gas-phase systems[9]). We term this ionization as 'unphysical', since such an effect would not occur in the bulk liquid. To test this, we perform similar calculations at a higher laser power to compensate for the renormalization of the KS potential. Figure S2(b) shows that indeed the full TDDFT calculation at a slightly higher laser power is nearly equivalent to the non-interacting calculation at a lower laser power, indicating that the source of the disparity is a consequence of ionization. In the limit of an infinitely large cluster this effect would not take place; thus, the results overall validate the use of this approximation, at least as a starting point for exploring strong-field physics in liquids.

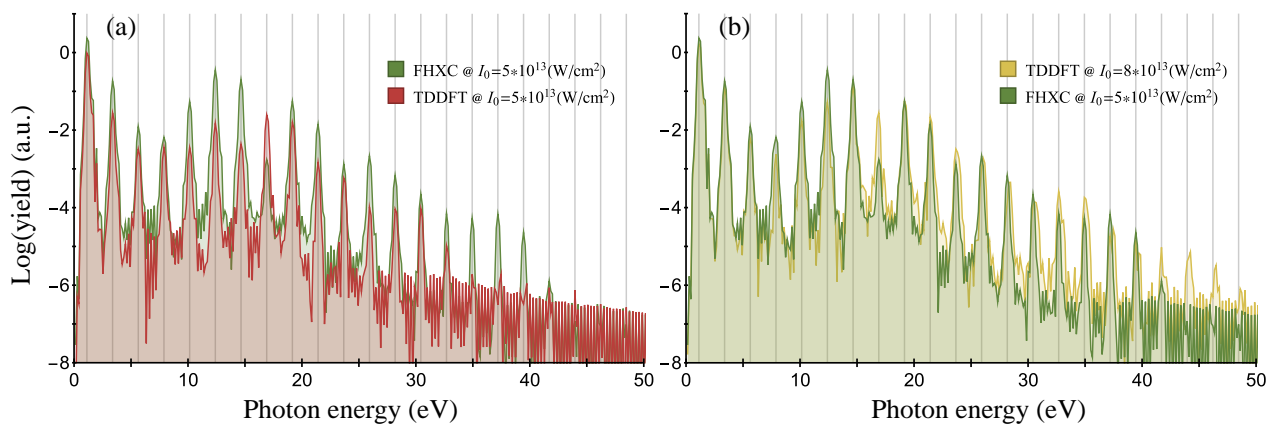

**Figure S2.** HHG spectra from liquid methane with and without the frozen KS potential approximation. (a) HHG spectra obtained at $\lambda=1100$nm and a laser power of $5\times10^{13}$ W/cm$^2$. (b) The same as in (a), but where the full TDDFT calculations are performed at a laser power of $8\times10^{13}$ W/cm$^2$ to compensate for the renormalization of the KS potential due to unphysical ionization. Gray lines indicate the position of odd harmonics.

## 2. Validity of frozen deep-lying states approximation

We demonstrate here the validity of the frozen deep-states approximation used throughout calculations. Figure S3(a) and (b) present HHG spectra from liquid water and ammonia, respectively, where the deepest-lying band of states has been either kept frozen, or allowed to evolve in time (i.e. in the frozen case the lowest $N$ states are frozen where $N$ is the number of molecules in the cluster). Results clearly show that in both cases this approximation is accurate (some disparities arise near 15eV for ammonia, but at an even harmonic order that vanishes after full orientation averaging). Figure S3(c) presents a similar analysis for liquid methane, where it is shown that the response of perturbative (below band gap) harmonics can drastically change if the deepest orbital band of methane is not included in the dynamics. In fact, very large disparities arise for the perturbative harmonics (especially the 5$^{th}$ harmonic) even when only the first 10 states are frozen. Thus, this approximation is not used throughout the paper for methane, only for water and ammonia. The physical source of the disparity in methane remains to be investigated, but could result from polarization of inner states[10]. We emphasize that for higher order harmonics the approximation remains valid just as expected.

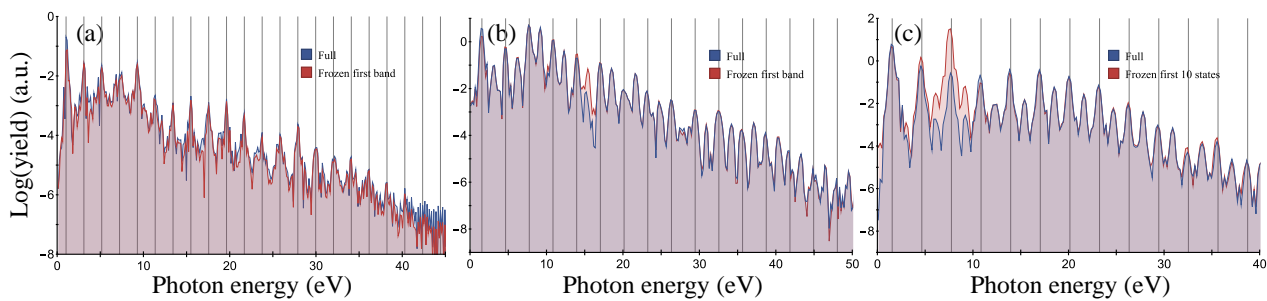

**Figure S3.** HHG spectra from various liquids with and without the frozen deep-lying states approximation. (a) HHG spectra from water obtained at $\lambda=1200$nm and a laser power of $4\times10^{13}$ W/cm$^2$. (b) HHG spectra from ammonia obtained at $\lambda=800$nm and a laser power of $5\times10^{13}$ W/cm$^2$. (c) HHG spectra from methane obtained at $\lambda=800$nm and a laser power of $5\times10^{13}$ W/cm$^2$. Note that for methane and ammonia the spectra only include a single orientation without averaging, such that even harmonics are also observed. Gray lines indicate the position of odd harmonics.

- **REFERENCES**